\documentclass[12pt,british]{elsarticle}
\usepackage[T1]{fontenc}
\usepackage[latin9]{inputenc}
\usepackage{geometry}
\usepackage[active]{srcltx}
\usepackage{units}
\usepackage{amsbsy}
\usepackage{amssymb}
\usepackage{graphicx}
\usepackage{esint}

\makeatletter
\def\vec#1{\boldsymbol{#1}}
\def\d{\mathrm d}
\def\i{\mathrm i}

\makeatother

\usepackage{babel}
\begin{document}

\begin{frontmatter}{}

\title{Concept of a next-generation electromagnetic phase-shift flowmeter
for liquid metals}

\author{Richard Looney}

\ead{looneyr@uni.coventry.ac.uk }

\author{J\={a}nis Priede}

\ead{j.priede@coventry.ac.uk}

\address{Flow Measurement Research Centre, Coventry University, UK}
\begin{abstract}
We present a concept of an electromagnetic phase-shift flowmeter that
has a significantly reduced sensitivity to the variation of the electrical
conductivity of a liquid metal. A simple theoretical model of the
flowmeter is considered where the flow is approximated by a solid
finite-thickness conducting layer which moves in the presence of an
ac magnetic field. In contrast to the original design {[}Priede et
al., Meas. Sci. Technol. 22 (2011) 055402{]}, where the flow rate
is determined by measuring only the phase shift between the voltages
induced in two receiving coils, the improved design measures also
the phase shift between the sending and the upstream receiving coils.
These two phase shifts are referred to as internal and external ones,
respectively. We show that the effect of electrical conductivity on
the internal phase shift, which is induced by the flow, can be strongly
reduced by rescaling it with the external phase shift, which depends
mostly on the conductivity of medium. Two different rescalings are
found depending on the ac frequency. At low frequencies, when the
shielding effect is weak, the effect of conductivity is strongly reduced
by rescaling the internal phase shift with the external one squared.
At higher frequencies, the same is achieved by rescaling the internal
phase shift directly with the external one. 
\end{abstract}

\end{frontmatter}{}

\section{Introduction}

Electromagnetic flow metering based on the voltage induced between
the electrodes in contact with a conductive liquid that flows in the
presence of magnetic field is a well established technique which works
reliably for common liquids with conductivities as low as that of
tap water, i.e., a few $\unit{\mu S/cm}$ \citep{Baker2016}. Application
of this standard technique to liquid metals is however hampered by
various contact problems ranging from the corrosion of electrodes
to spurious electrochemical and thermoelectric potentials \citep{Schulenberg2010}.
The effect of spurious potentials can be mitigated by using a pulsed
magnetic field \citep{Velt2013}. The corrosion problem could in principle
be avoided by using capacitately-coupled electrodes to measure the
induced voltage in a contactless way \citep{Hussain1985,McHale1985}. 

There are several more widely used contactless flow metering methods
for liquid metals which rely on the eddy currents induced by the flow.
As first suggested by \citet{Shercliff1987} and later pursued by
the so-called Lorentz Force Velocimetry \citep{Thess2007a}, the flow
rate can be determined by measuring the force generated by eddy currents
on a magnet placed close to the pipe carrying a conducting liquid.
An alternative approach, which is virtually force-free and thus largely
independent of the conductivity of liquid metal \citep{Priede2009b},
is to determine the flow rate from the equilibrium rotation rate of
a freely rotating magnetic flywheel \citep{Shercliff1960,Bucenieks2005,Hvasta2018}
or just a single magnet \citep{Priede2011b}. More common are the
contactless eddy-current flow meters which measure the the flow-induced
perturbation of an externally applied magnetic field \citep{Lehde1948,Cowley1965,Poornapushpakala2010a}.
The same principle is used also by the so-called flow tomography approach,
where the spatial distribution of the induced magnetic field is analysed
\citep{Stefani2004,Stefani2000}. 

\global\long\def\Rm{\mathit{Rm}}
 The main challenge to the eddy current flowmetering is the weak induced
magnetic field with a relative amplitude of the order of magnitude
of the magnetic Reynolds number $\Rm\sim10^{-4}-10^{-1}$, which has
to be measured in the presence of a much stronger external magnetic
field. Although the background signal produced by the external magnetic
field can be compensated by a proper arrangement of sending and receiving
coils \citep{Feng1975}, this type of flow meter remains highly susceptible
to small geometrical imperfections. 

Much more robust to geometrical disturbances is the phase-shift induction
flowmeter \citep{Baker2016,Priede2011c,Priede2012a}. Instead of the
usual voltage difference, this flowmeter measures the phase shift
induced by the flow between the voltages in two receiver coils. Because
the phase is determined by the amplitude ratio, it remains invariant
when the amplitudes are perturbed in a similar way. The main remaining
drawback of this flowmeter is the dependence of the flow-induced phase
shift not only on the velocity but also on the conductivity of liquid
metal. 

The susceptibility to varying conductivity is a general problem which
plagues not only eddy-current \citep{Poornapushpakala2014} but also
the Lorentz force devices. While there are devices which can address
this problem, such as, for example, the single magnet rotary flowmeter
\citep{Priede2011b}, the transient eddy-current flowmeter \citep{Forbriger2015,Krauter2017a,Looney2018a}
or the time-of-flight Lorentz force flowmeters \citep{Vire2010},
they come with other drawbacks such as moving parts or complicated
measurement schemes. For standard eddy-current flowmeters, the effect
of conductivity can be compensated by normalizing the voltage difference
between two sensing coils with the sum of the voltages \citep{Pavlinov2017}.
However, it has to be noted that this compensation scheme works only
at a certain optimal AC frequency which depends on the setup \citep{Sharma2010}.

This paper is concerned with the development of a next-generation
phase-shift flowmeter with a reduced dependence on the conductivity
of liquid metal. The basic idea is that not only the eddy currents
induced by the liquid flow but also those induced by the alternating
magnetic field itself give rise to a phase shift in the induced magnetic
field and the associated emf which is picked up by the sensor coils.
The latter effect, which leads to a phase shift between the sending
and receiving coils, depends on the conductivity of the liquid. Therefore,
it may be used to compensate for the effect of conductivity on the
flow-induced phase shift. The feasibility of such an approach is investigated
using two simple theoretical models of the phase-shift flowmeter,
where the flow is approximated by a solid body motion of a finite-thickness
conducting layer.

The paper is organised as follows. Mathematical model and the basic
equations describing the phase-shift flowmeter are briefly recalled
in the next section. In Section \ref{sec:Num} we present and discuss
numerical results for two simple configurations of the applied magnetic
field: the first in the form of a mono-harmonic standing wave and
the second generated by a simple coil made of couple of straight wires
placed above the layer. The paper is concluded by a summary in Section
\ref{sec:Sum}. Relevant mathematical details of solutions behind
numerical results are provided in the Appendix. 

\section{\label{sec:Math}Mathematical model}

\begin{figure}
\begin{centering}
\includegraphics[width=0.6\columnwidth]{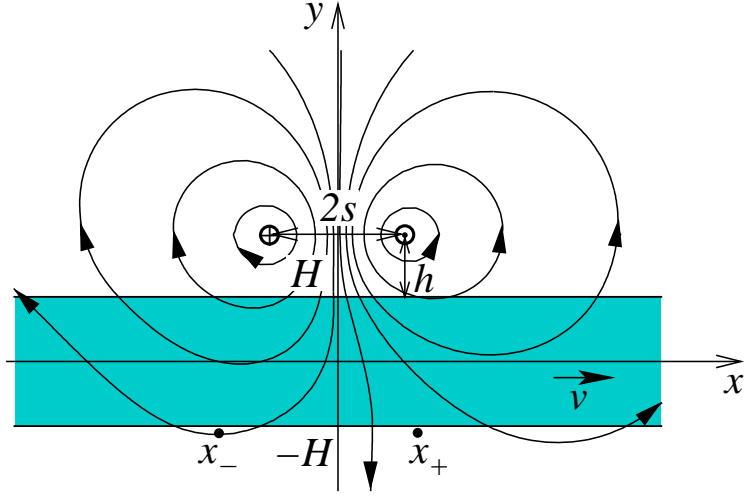}
\par\end{centering}
\caption{\label{cap:sketch}Sketch of the mathematical model of flowmeter consisting
of an electrically conducting layer which moves as a solid body in
the presence of alternating magnetic field generated by a couple of
straight wires with opposite currents. Flow rate is determined by
measuring the phase difference between the voltages induced in two
sensor wires (coils) placed directly beneath the layer at the downstream
and upstream horizontal positions $x_{-}$ and $x_{+},$ respectively. }
\end{figure}

Consider a layer of electrical conductivity $\sigma$ and thickness
$2H$ which moves as a solid body with the velocity $\vec{v}=\vec{e}_{x}v$
in the presence of an alternating magnetic field $\vec{B}$ as shown
in figure \ref{cap:sketch}. The velocity profile as well as other
technical details are ignored in this basic model whose main purpose
to capture the key characteristics of flowmeter. The electric field
induced by alternating magnetic field follows from the Maxwell-Faraday
equation $\vec{E}=-\vec{\nabla}\Phi-\partial_{t}\vec{A},$ where $\Phi$
is the electric potential and $\vec{A}$ is the magnetic vector potential
of the magnetic field $\vec{B}=\vec{\nabla}\times\vec{A}$. The electric
current induced in the moving medium is given by Ohm's law 
\[
\vec{j}=\sigma(\vec{E}+\vec{v}\times\vec{B})=\sigma(-\vec{\nabla}\Phi-\partial_{t}\vec{A}+\vec{v}\times\vec{\nabla}\times\vec{A}).
\]
 Subsequently, we consider a 2D magnetic field which is invariant
in the $z$ direction (perpendicular to the plane of figure \ref{cap:sketch})
and alternates in time harmonically with the angular frequency $\omega.$
Such a magnetic field, which has only two $(x$ and $y)$ components,
is described by single $(z)$ component of the vector potential: $\vec{A}=\vec{e}_{z}A.$
In this case, we have $\vec{B}=-\vec{e}_{z}\times\vec{\nabla}A,$
which means $\vec{B}\cdot\vec{\nabla}A\equiv0,$ i.e., the isolines
of $A$ are the flux lines of $\vec{B}.$ A harmonically alternating
solution can be found in the complex form as $A(\vec{r},t)=\Re\left[A(\vec{r})e^{\i\omega t}\right]$,
where $A(\vec{r})$ is an amplitude distribution and $\vec{r}$ is
the radius vector. 

At sufficiently low alternation frequencies, which are typical for
eddy current flowmeters, the displacement currents are negligible,
and thus the Ampere's law $\vec{j}=\mu_{0}^{-1}\vec{\nabla}\times\vec{B}$
leads to $\Phi\equiv0.$ As a result, we obtain the following advection-diffusion
equation for $A(\vec{r})$

\begin{equation}
\mu_{0}\sigma(\i\omega A+v\partial_{x}A)=\vec{\nabla}^{2}A,\label{eq:A}
\end{equation}
where $\mu_{0}=\unit[4\pi\times10^{-7}]{H/m}$ is the vacuum permeability.
The continuity of $\vec{B}$ at the interface $S$ between the conducting
medium and free space requires 
\begin{equation}
\left[A\right]_{S}=\left[(\vec{n}\cdot\vec{\nabla})A\right]_{S}=0,\label{bc:A}
\end{equation}
 where $\left[\right]_{S}$ and $\vec{n}$ respectively denote a jump
across $S$ and the unit normal to $S.$

In the following, we consider two basic configurations of the applied
magnetic field \citep{Priede2011c}. The first is a mono-harmonic
standing wave with the wave number $k$ in the $x$-direction
\begin{equation}
A_{0}(\boldsymbol{r},t)=\hat{A}_{0}(y)\cos(kx)\cos(\omega t).\label{eq:mono}
\end{equation}
The second is a magnetic field generated by an idealized finite-size
coil which consists of two parallel straight wires carrying an ac
current of amplitude $I_{0}$ that flows in opposite directions along
the $z$-axis at distance $2s$ in the $x$-direction and the height
$h$ above the upper surface of the layer, as shown in figure \ref{cap:sketch}. 

The two key parameters which appear in the problem are the dimensionless
ac frequency $\bar{\omega}=\mu_{0}\sigma\omega H^{2}$ and the magnetic
Reynolds number $\Rm=\mu_{0}\sigma vH.$ The latter defines the velocity
of layer in the units of the characteristic magnetic diffusion speed
$(\mu_{0}\sigma H)^{-1}$ and is subsequently referred to as the dimensionless
velocity. It is important to note that the dimensionless velocity
defined in this way varies with the conductivity of medium. This variation
is eliminated by taking the ratio $\Rm/\bar{\omega}=v/(\omega H)\equiv\bar{v},$
which defines the velocity of layer relative to the characteristic
phase speed of the magnetic field and is referred to as the relative
velocity. For fixed $H$ and $\omega,$ the relative velocity depends
only on the physical velocity $v,$ while $\bar{\omega}$ depends
only on the conductivity of medium. Consequently, for the flowmeter
to be insensitive to the conductivity variation of medium, its output
signal should depend only on $\bar{v}$ but not on $\bar{\omega.}$ 

\section{\label{sec:Num}Results}

\subsection{Mono-harmonic magnetic field}

\begin{figure}
\begin{centering}
\includegraphics[width=0.6\columnwidth]{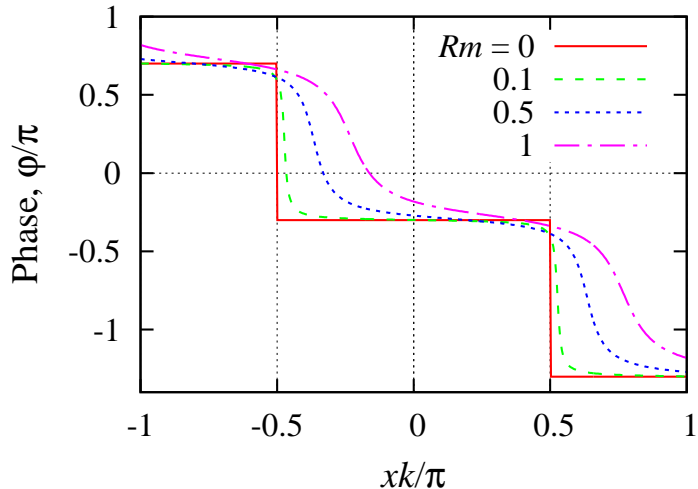}
\par\end{centering}
\caption{\label{cap:ph2-x}Phase distribution of the vector potential over
a wavelength of the applied magnetic field along the bottom of the
layer for $\bar{\omega}=1,$ $k=1$ and various dimensionless velocities
$\protect\Rm.$}
\end{figure}

The problem formulated above has an analytical solution \citep{Priede2011c}
which is used in the following. To make the paper self-contained,
a summary of solution is provided in the Appendix.

We start with the external magnetic field in the form a mono-harmonic
standing wave (\ref{eq:mono}) and first briefly recall the basic
characteristics of the original phase-shift flowmeter \citep{Priede2011c}.
When the layer is at rest $(\Rm=0),$ the induced magnetic field forms
a standing wave like the external field. As the adjacent halves of
standing wave oscillate in opposite phases, the phase distribution
is piece-wise continuous along the wave. The phase of vector potential
jumps by $\pi$ across the wave nodes, which for the $k=1$ shown
in figure \ref{cap:ph2-x} are located at $x=\pm\pi/2.$ When the
layer starts to move, the phase discontinuities are smoothed out and,
with the increase of velocity, are shifted further downstream. As
seen in figure \ref{cap:ph2-x}, the strongest phase variation produced
by the motion occurs just downstream of the wave nodes whilst the
variation upstream remains relatively weak.

For the physical interpretation of these and subsequent results, note
that the circulation of vector potential along a closed contour gives
a magnetic flux through the encircled surface. For the 2D case under
consideration, the vector potential has only one $(z)$ component
$A$, and the difference of $A$ between two points defines the linear
flux density between the two lines parallel to the vector potential
at those two points. The same holds also for the time derivative of
the corresponding quantities. Thus, the difference of the vector potential
amplitudes between two points is proportional to the voltage amplitude
measured by an idealised coil consisting of two straight parallel
wires placed along the $z$-axis at those points. Correspondingly,
the single-point vector potential gives the voltage measured by a
`wide' coil with the second wire placed sufficiently far away in the
region of a negligible magnetic field.

\begin{figure}
\begin{centering}
\includegraphics[width=0.5\columnwidth]{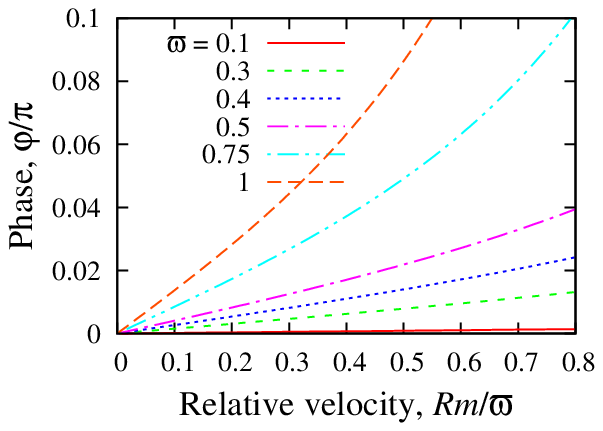}\put(-35,40){(\textit{a})}\includegraphics[width=0.5\columnwidth]{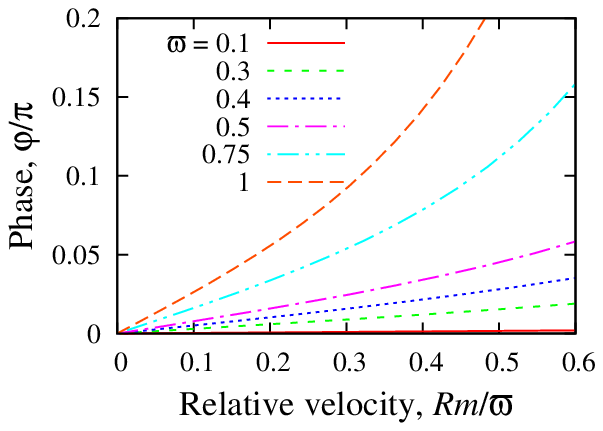}\put(-35,40){(\textit{b})}
\par\end{centering}
\begin{centering}
\includegraphics[width=0.5\columnwidth]{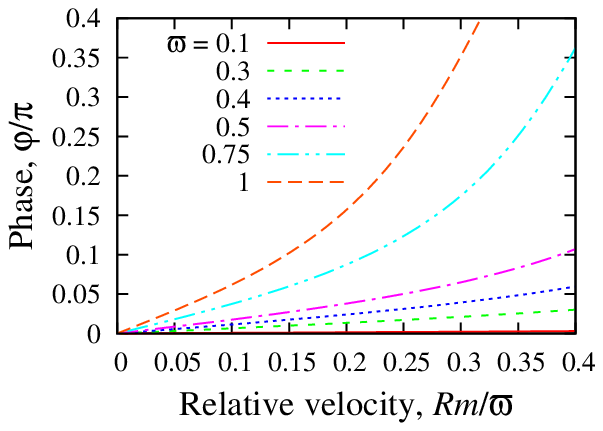}\put(-35,40){(\textit{c})}
\par\end{centering}
\caption{\label{cap:ph2-dif}The phase shift $\Delta_{0}\varphi$ (scaled by
$\pi)$ between two observation points placed below the layer at $\pm xk/\pi=0.2$
(a), $0.3$ (b) and $0.4$ (c) versus the relative velocity $\protect\Rm/\bar{\omega}$
for $k=1$ at various dimensionless frequencies $\bar{\omega}\lesssim1.$ }
\end{figure}

In the original concept, the flow rate is determined by measuring
only the phase shift $\Delta_{0}\varphi$ between two observation
points placed symmetrically with respect $x=0,$ which is the midpoint
between two adjacent nodes of the mono-harmonic standing wave (\ref{eq:mono}).
The variation of this phase shift with the relative velocity $\bar{v}=v/(\omega H)$
is shown in figure \ref{cap:ph2-dif} for various dimensionless ac
frequencies $\bar{\omega},$ which correspond to various conductivities
when the physical ac frequency $\omega$ is kept fixed. Since $\bar{v}$
is independent of the conductivity, the variation of $\Delta_{0}\varphi$
with $\bar{\omega}$ for a fixed $\omega$ is entirely due to the
variation of conductivity. Since the flow-induced phase shift vanishes
at low velocities, and according to Eq. (\ref{eq:A}) it does so directly
with $\Rm,$ we have $\Delta_{0}\varphi\sim\Rm$ for $\Rm\ll1.$ By
the same arguments, for low ac frequencies, we have $\Delta_{0}\varphi\sim\bar{\omega.}$
For $\Rm\ll1$ and $\bar{\omega}\ll1,$ the combination of these two
relations imply a quadratic variation of the phase shift with the
electrical conductivity of the medium:
\begin{equation}
\Delta_{0}\varphi\sim\bar{\omega}\Rm\sim\sigma^{2}.\label{eq:phi0}
\end{equation}

Our goal is to compensate the variation of the flow-induced phase
shift with the conductivity by employing the information contained
in the phase shift between the sending coil and one of the receiving
coils. Namely, we shall use the phase shift between the sending and
the upstream receiving coil, which is less affected by the motion
of the conducting medium than the downstream one, to rescale the flow-induced
phase shift between the voltages in the receiving coils.

\begin{figure}
\begin{centering}
\includegraphics[width=0.5\columnwidth]{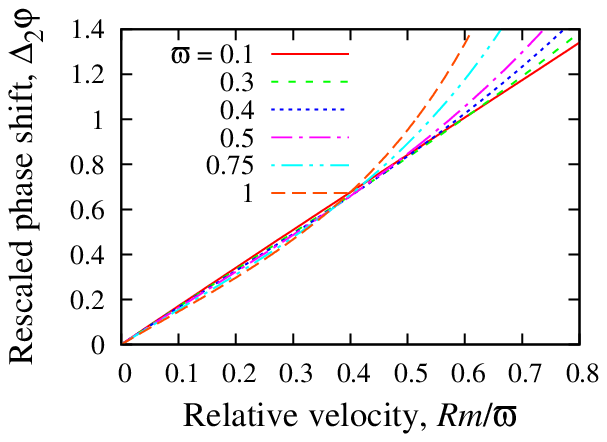}\put(-35,40){(\textit{a})}\includegraphics[width=0.5\columnwidth]{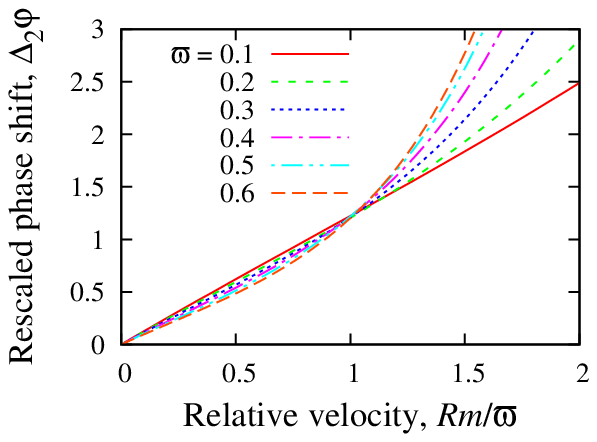}\put(-35,40){(\textit{b})}
\par\end{centering}
\begin{centering}
\includegraphics[width=0.5\columnwidth]{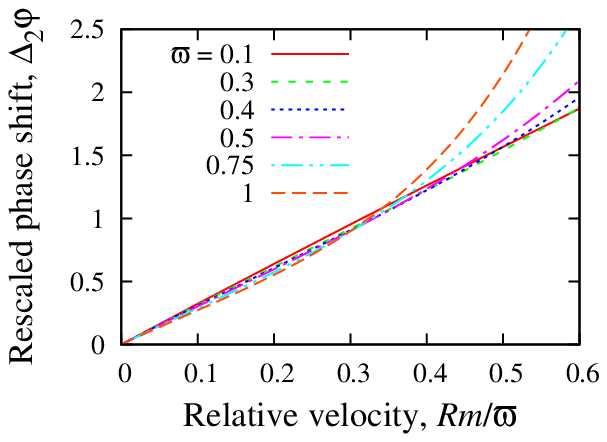}\put(-35,40){(\textit{c})}\includegraphics[width=0.5\columnwidth]{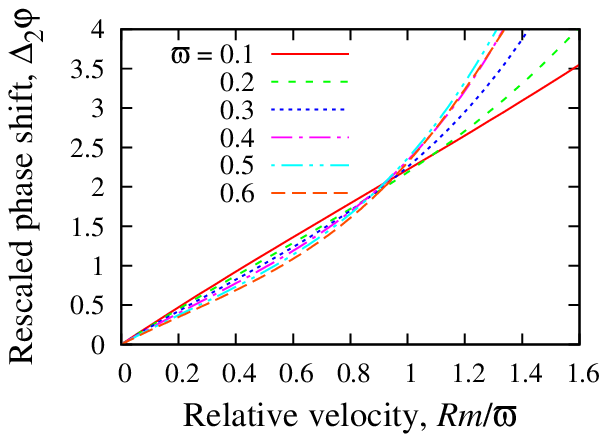}\put(-35,40){(\textit{d})}
\par\end{centering}
\begin{centering}
\includegraphics[width=0.5\columnwidth]{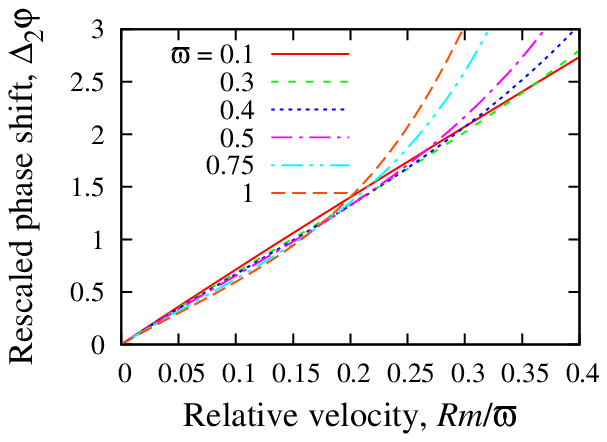}\put(-35,40){(\textit{e})}\includegraphics[width=0.5\columnwidth]{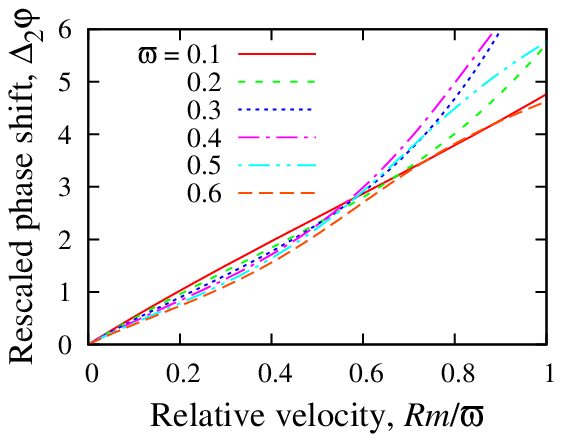}\put(-35,40){(\textit{f})}
\par\end{centering}
\caption{\label{cap:ph2a-dif}Rescaled phase shift $\Delta_{2}\varphi$ between
two observation points placed below the layer at $\pm xk/\pi=0.2$
(a,b), $0.3$ (c,d) and $0.4$ (e,f) versus the relative velocity
$\protect\Rm/\bar{\omega}$ for $k=1$ (a,c,e) and $k=0.5$ (b,d,f)
at various dimensionless frequencies $\bar{\omega}\lesssim1.$ }
\end{figure}

Let us start with a low dimensionless frequency $\bar{\omega}\ll1.$
In this case, the phase shift between sending and receiving coils,
which we refer to as the external phase shift, and which is caused
by the diffusion of the magnetic field through the conducting layer
is expected to vary as $\varphi\sim\bar{\omega}$. This variation,
which follows from the same arguments as Eq. (\ref{eq:phi0}), implies
that the conductivity can be eliminated from the flow-induced phase
shift (\ref{eq:phi0}) by the following rescaling
\begin{equation}
\Delta_{2}\varphi=\frac{\Delta_{0}\varphi}{\varphi_{-}^{2}},\label{eq:dhi2}
\end{equation}
where $\Delta_{0}\varphi=\varphi_{+}-\varphi_{-}$ is a difference
between the phases of voltage, $\varphi_{+}$ and $\varphi_{-},$
measured at downstream and upstream observation points. For a fixed
physical ac frequency, the variation of the dimensionless frequency
$\bar{\omega}$ can be only due to by the variation of electrical
conductivity which would cause also a variation of the dimensionless
velocity $\Rm$ independently of the variation of physical velocity.
As discussed above, the quantity which varies with the physical velocity
but not with the conductivity is the relative velocity $\bar{v}=\Rm/\bar{\omega}.$
It means that in the ideal case of a perfectly compensated conductivity
effect, the rescaled phase shift $\Delta_{2}\varphi$ should depend
only on $\bar{v}$ but on $\bar{\omega.}$ Practically, we expect
the rescaled phase shift to vary mostly with $\bar{v}$ and not so
much with $\bar{\omega.}$ This can be seen to be the case for the
rescaled phase shifts plotted in figure \ref{cap:ph2a-dif} which
show a much weaker dependence on the dimensionless frequency than
the unscaled phase shift shown in figure \ref{cap:ph2-dif}. For low
relative velocities, the variation of rescaled phase shift with $\bar{\omega}$
remains rather weak up to $\bar{\omega}\sim1,$ which is consistent
with the original assumption of $\bar{\omega}\ll1.$ The range of
relative velocities which is weakly affected by $\bar{\omega}$ depends
on the location of the observation points. The closer the observation
points to the wave nodes $(xk/\pi=\pm0.5),$ the shorter the range
of the relative velocities for which $\Delta_{2}\varphi$ remains
insensitive to the variation of $\bar{\omega}$. In this range of
relative velocities, the increase of the dimensionless frequency from
$0.1$ to $1,$ which is equivalent to the increase of the conductivity
by an order of magnitude, results in a much smaller variation of the
rescaled phase shift.

\begin{figure}
\begin{centering}
\includegraphics[width=0.5\columnwidth]{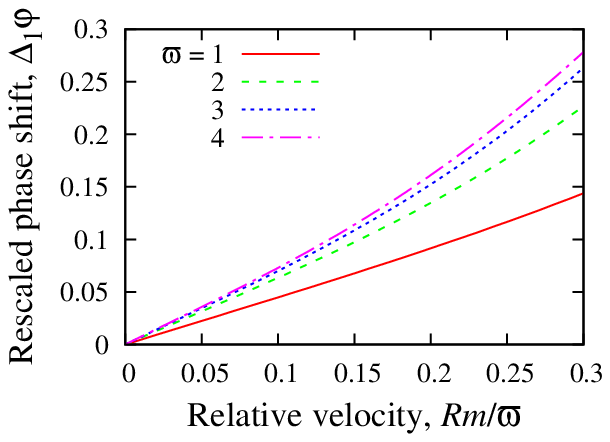}\put(-35,40){(\textit{a})}\includegraphics[width=0.5\columnwidth]{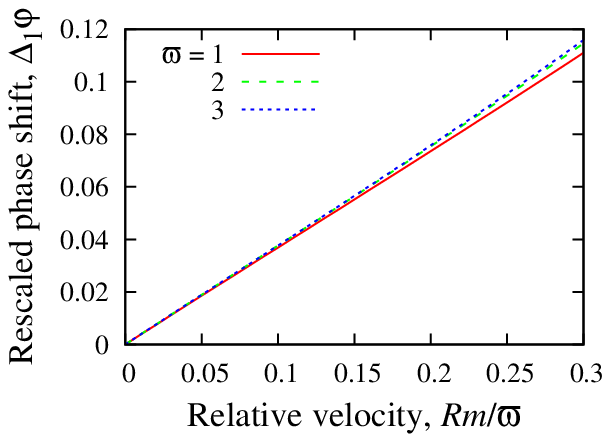}\put(-35,40){(\textit{b})}
\par\end{centering}
\begin{centering}
\includegraphics[width=0.5\columnwidth]{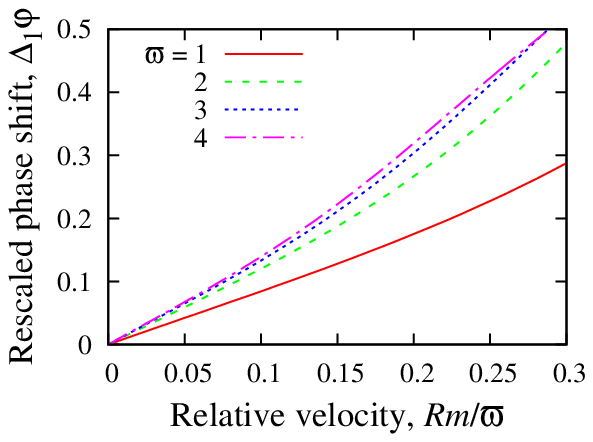}\put(-35,40){(\textit{c})}\includegraphics[width=0.5\columnwidth]{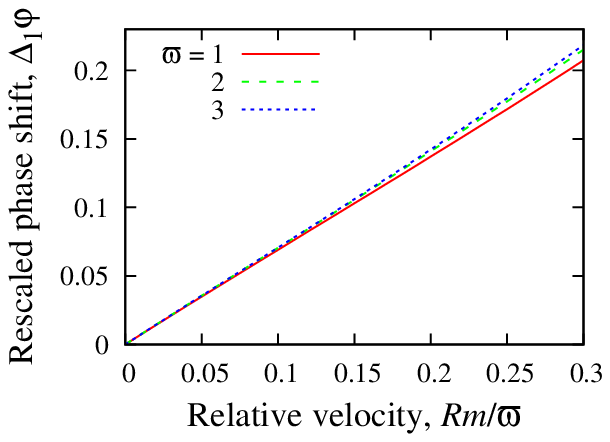}\put(-35,40){(\textit{d})}
\par\end{centering}
\begin{centering}
\includegraphics[width=0.5\columnwidth]{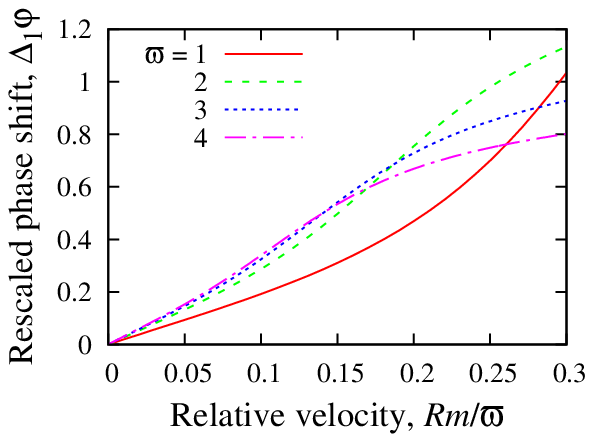}\put(-35,40){(\textit{e})}\includegraphics[width=0.5\columnwidth]{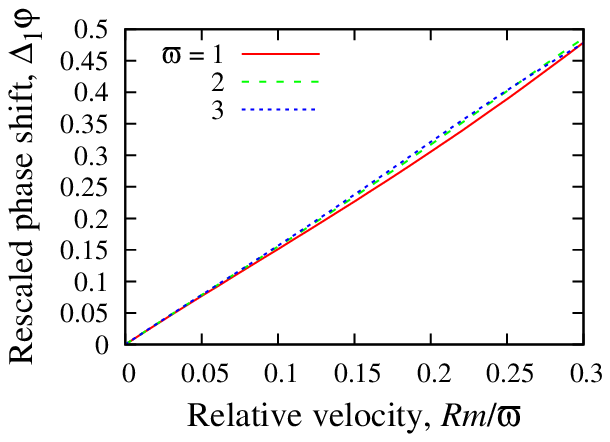}\put(-35,40){(\textit{f})}
\par\end{centering}
\caption{\label{cap:ph2b-dif}Rescaled phase shift $\Delta_{1}\varphi$ between
two observation points placed below the layer at $\pm xk/\pi=0.2$
(a,b), $0.3$ (c,d) and $0.4$ (e,f) versus the relative velocity
$\protect\Rm/\bar{\omega}$ for $k=1$ (a,c,e) and $k=0.5$ (b,d,f)
at various dimensionless frequencies $\bar{\omega}\gtrsim1.$ }
\end{figure}

The above results show that rescaling (\ref{eq:dhi2}) can compensate
the effect of conductivity only at sufficiently low frequencies at
which the variation of the frequency-induced (external) phase shift
between the sending and receiving coils is close to linear with $\bar{\omega}.$
This scaling changes at higher frequencies when the shielding effect
makes the variation of this phase shift non-linear with $\bar{\omega.}$
In this case, a new scaling can be deduced by the following order-of-magnitude
arguments which will be confirmed by the subsequent numerical results. 

For $\bar{\omega}\gg1,$ a significant phase variation of order $\sim1$
occurs in the conducting medium over the the characteristic skin layer
thickness $\sim\bar{\omega}^{-1/2}.$ Consequently, the total phase
shift due to the diffusion of the magnetic field through the whole
conducting layer with the dimensionless thickness $\sim1$ is expected
to vary as $\varphi\sim\bar{\omega}^{1/2}\sim\sigma^{1/2}.$ On the
other hand, since the applied magnetic field in the form of a standing
wave consists of two oppositely travelling waves (see , Eq. (\ref{sol:A-i})
suggests that the motion of the layer is equivalent to the variation
of the dimensionless frequency $\bar{\omega}$ by $k\Rm\sim\Rm\ll1.$
Then the flow-induced phase shift between two receiving coils is expected
to scale as 
\[
\Delta_{0}\varphi\sim\frac{\partial\varphi}{\partial\bar{\omega}}\Rm\sim\bar{\omega}^{-1/2}\Rm\sim\sigma^{1/2}.
\]
This implies that for $\bar{\omega}\gtrsim1$ the conductivity can
be eliminated from the flow-induced phase shift between the receiving
coils by rescaling it directly with the external phase shift as
\begin{equation}
\Delta_{1}\varphi=\frac{\Delta_{0}\varphi}{\varphi_{-}}.\label{eq:dhi1}
\end{equation}

The rescaled phase shift $\Delta_{1}\varphi$ is plotted in figure
\ref{cap:ph2b-dif} against the relative velocity for two different
wavenumbers $k=1$ and and $k=0.5$ and various dimensionless frequencies.
For $k=1,$ the dependence of $\Delta_{1}\varphi$ on $\bar{\omega}$
is seen to diminish as the latter is increased above $1.$ For $k=0.5,$
which corresponds a wavelength of the applied magnetic field significantly
larger than the thickness of the layer, the variation of $\Delta_{1}\varphi$
with $\bar{\omega}$ is practically insignificant starting from $\bar{\omega}\approx1.$ 

\subsection{External magnetic field generated by a pair of straight wires}

\begin{figure*}
\begin{centering}
\includegraphics[width=0.6\columnwidth]{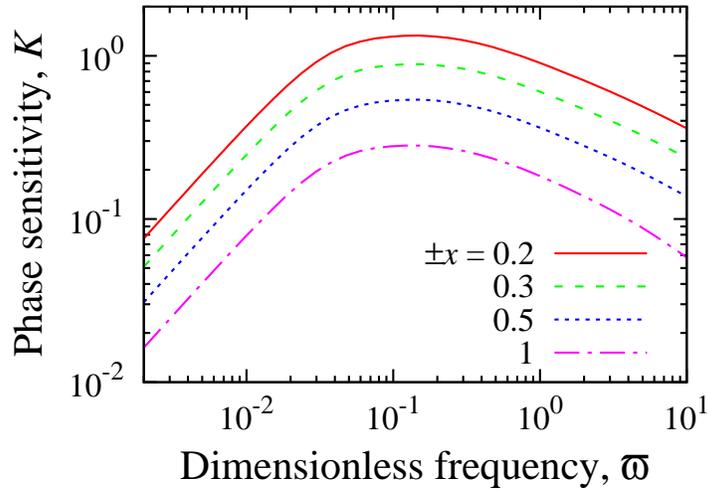}
\par\end{centering}
\caption{\label{cap:sens} The phase sensitivity $K=\frac{1}{\pi}\left.\frac{\partial\varphi}{\partial\protect\Rm}\right|_{\protect\Rm=0}$
versus the dimensionless frequency $\bar{\omega}$ at various observation
points at the bottom of the layer .}
\end{figure*}

\begin{figure}
\begin{centering}
\includegraphics[width=0.5\columnwidth]{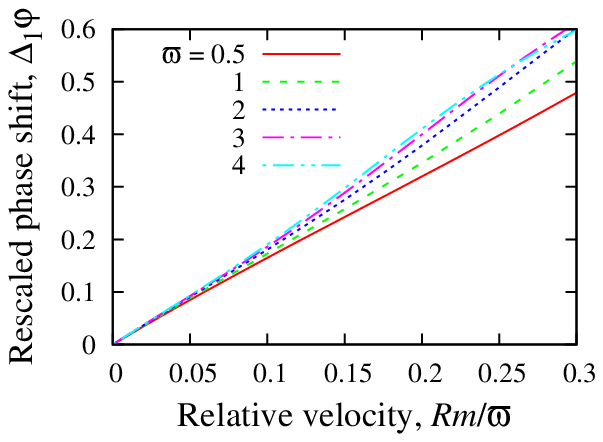}\put(-35,40){(\textit{a})}\includegraphics[width=0.5\columnwidth]{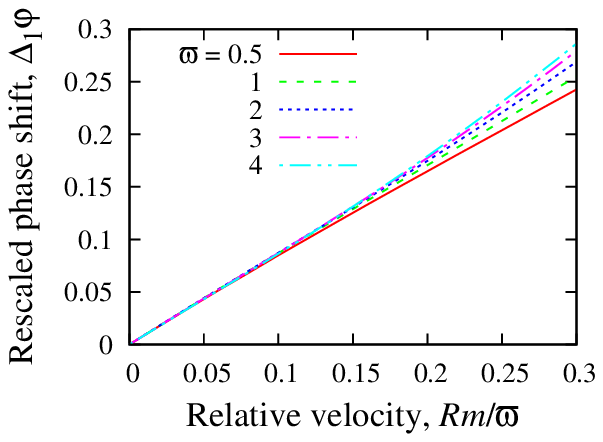}\put(-35,40){(\textit{b})}
\par\end{centering}
\begin{centering}
\includegraphics[width=0.5\columnwidth]{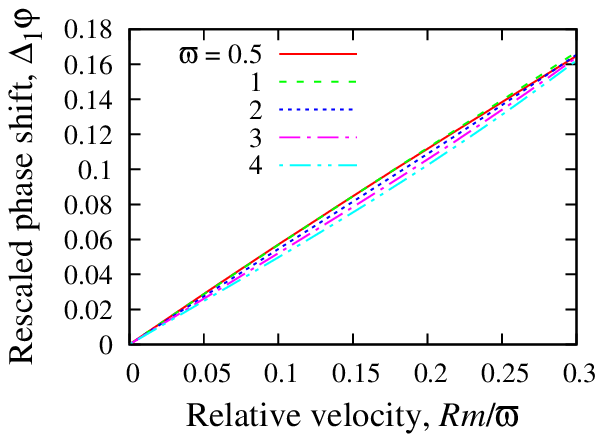}\put(-35,40){(\textit{c})}\includegraphics[width=0.5\columnwidth]{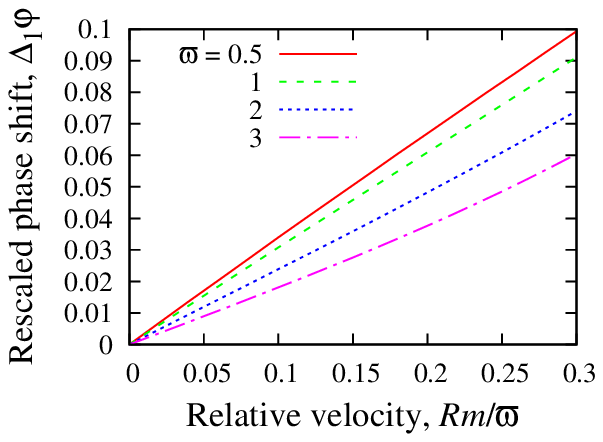}\put(-35,40){(\textit{d})}
\par\end{centering}
\caption{\label{cap:ph3r-dif}The rescaled phase shift $\Delta_{1}\varphi$
between two observation points placed below the layer at $\pm x=0.5$
(a), $1$ (b), $1.5$ (c), and $2.5$ (d) versus the relative velocity
$\protect\Rm/\bar{\omega}$ at various dimensionless frequencies $\bar{\omega}.$ }
\end{figure}

In this section, we consider a more realistic external magnetic field
which is generated by a couple of parallel wires with opposite currents.
The wires are separated by a horizontal distance of $2s=2$ and put
at the height $h=1$ above the layer as shown in figure \ref{cap:sketch}.
For a layer at rest, the magnetic field distribution is mirror-symmetric
with respect to the $x=0$ plane, which is analogous to a node in
the mono-harmonic standing wave considered in the previous section.
Correspondingly, when the layer is at rest, there is a phase jump
of $\pi$ at $x=0.$ In contrast to the previous case, the phase is
no longer constant on either side of the discontinuity and varies
horizontally as well as vertically \citep{Priede2011c}. As a result,
the range of dimensionless frequency where the phase sensitivity $K=\frac{1}{\pi}\left.\frac{\partial\varphi}{\partial\Rm}\right|_{\Rm=0},$
which is plotted in figure \ref{cap:sens}, varies more or less linearly
with $\bar{\omega}$ and, thus, rescaling (\ref{eq:dhi2}) could be
applicable, is relatively short and limited to $\bar{\omega}\lesssim0.02.$
Therefore, at $\bar{\omega}\sim1,$ which presents the main interest
from a practical point of view, scaling (\ref{eq:dhi1}) is expected
to be applicable. The phase shift rescaled in this way is plotted
in figure \ref{cap:ph3r-dif} against the relative velocity for several
locations of the observation points and various dimensionless frequencies.
As seen, $\Delta_{1}\varphi$ depends essentially on the relative
velocity while its variation with $\bar{\omega}$ is relatively weak
except for the largest separation of the observation points $x=\pm2.5,$
which is shown in figure \ref{cap:ph3r-dif}(d). This deterioration
of the rescaling at larger separations of observation points is likely
due to the horizontal variation of the external phase shift mentioned
above.

\section{\label{sec:Sum}Summary and conclusion}

We have presented a concept of an improved phase-shift flowmeter which
is much less susceptible to the variations of electrical conductivity
of liquid metal than the original design of \citet{Priede2011c}.
In the improved flowmeter, we suggest to measure not only the phase
shift between the voltages induced in two receiving coils, which is
referred to as the internal phase shift, but also the external phase
shift between the sending coil and the upstream receiving coil. In
contrast to the internal phase shift, which is induced by the flow,
the external one depends mostly on the conductivity of the media and
not so much on its velocity. We rescale the internal phase shift with
the external phase shift to eliminate, or at least strongly reduce,
the effect of conductivity on the operation of the flowmeter. Two
different rescalings are found depending on the ac frequency. At low
frequencies $\bar{\omega}\lesssim1,$ when the phase shift varies
directly with the frequency, the conductivity can be eliminated by
rescaling the internal phase shift with the square of the external
phase shift. At higher ac frequencies $\bar{\omega}\gtrsim1,$ where
the shielding effect makes the variation of phase with the frequency
non-linear, the conductivity can be eliminated by rescaling the internal
phase shift directly with the external one. Note that for liquid sodium
with $\sigma=8.3\times10^{6}\,\mbox{S/m}$ \citep{Mueller2001} and
characteristic size $H\sim\unit[0.1]{m,}$ $\bar{\omega}\sim1$ and
$\Rm\sim1$ correspond to ac frequency $f\sim\unit[60]{Hz}$ and velocity
$v\sim\unit[1]{m/s,}$ respectively. 

The applicability of the first rescaling is limited to relatively
low frequencies $(\bar{\omega}\ll1)$, especially for realistic sending
coils which generate the magnetic field dominated by long-wave harmonics.
A potential disadvantage of using low ac frequencies may be the relatively
low sensitivity of the phase-shift flowmeter. From this point of view,
it seems more attractive to operate the flowmeter in the frequency
range with a moderate shielding effect, i.e. with $\bar{\omega}\sim1,$
where the second (direct) rescaling is applicable. Although rescaling
can render the phase-shift flowmeter virtually insensitive to the
conductivity variation of liquid metal, the flowmeter still requires
calibration because the output signal is not related straightforwardly
with the flow velocity as in the transient eddy- current \citep{Forbriger2015,Krauter2017a,Looney2018a}
and the time-of-flight Lorentz force \citep{Vire2010} flowmeters.
The proposed concept can be used for designing a next-generation phase-shift
flowmeter with increased robustness to the variations of the electrical
conductivity of liquid metal, which may be required in some metallurgical
and other applications.

\section{\label{sec:App}Appendix}

\subsection{Analytical solution for mono-harmonic external magnetic field}

The amplitude distribution of the external magnetic (\ref{eq:mono})
in the free space, which follows from equation (\ref{eq:A}) with
$\sigma=0$, can be written as 
\begin{equation}
\hat{A}_{0}(y;k)=C_{0}e^{|k|(y-1)},\label{sol:A0}
\end{equation}
 where the constant 
\begin{equation}
C_{0}=\hat{A}_{0}(1;k)\label{eq:C0}
\end{equation}
defines the amplitude of the Fourier mode with the wavenumber $k$
at the upper boundary of the layer. Representing the external magnetic
field (\ref{eq:mono}) as a superposition of two oppositely travelling
waves 
\begin{equation}
A_{0}(\vec{r},t)=\frac{1}{2}\left[A_{0}^{+}(\vec{r},t)+A_{0}^{-}(\vec{r},t)\right],\label{eq:Apm}
\end{equation}
 where $A_{0}^{\pm}(\vec{r},t)=\hat{A}_{0}(y)\cos(\omega t\pm kx),$
we can write the general solution in a similar form as 
\[
A(\vec{r},t)=\frac{1}{2}\left[A^{+}(\vec{r},t)+A^{-}(\vec{r},t)\right],
\]
where $A^{\pm}(\vec{r},t)=\Re\left[\hat{A}(y;\pm k)e^{\i(\omega t\pm kx)}\right].$
The solution in the free space above the layer $(y\ge1)$ is
\begin{equation}
\hat{A}(y;k)=\hat{A}_{0}(y;k)+\hat{A}_{1}(y;k),\label{sol:A-a}
\end{equation}
where the first term represents the external field (\ref{sol:A0})
and $\hat{A}_{1}(y;k)=C_{1}e^{-|k|(y-1)}$ is the induced field. In
the free space below the layer $(y\le-1)$, we have 
\begin{equation}
\hat{A}(y;k)=C_{3}e^{|k|(y+1)}.\label{sol:A-b}
\end{equation}
 In the conducting layer $(-1\le y\le1)$, the solution can be written
as 
\begin{equation}
\hat{A}(y;k)=C_{2}\sinh(\kappa y)+D_{2}\cosh(\kappa y),\label{sol:A-i}
\end{equation}
where $\kappa=\sqrt{k^{2}+\i(\bar{\omega}+k\Rm)}$, $\bar{\omega}=\mu_{0}\sigma\omega H^{2}$
is a dimensionless ac frequency and $\Rm=\mu_{0}\sigma vH$ is the
magnetic Reynolds number, which represents a dimensionless velocity.
The unknown constants, which depend on the wavenumber $k,$ are found
using the boundary conditions (\ref{bc:A}) as follows
\begin{eqnarray}
C_{2} & = & C_{0}|k|/\left(|k|\sinh(\kappa)+\kappa\cosh(\kappa)\right)\label{eq:C2}\\
D_{2} & = & C_{0}|k|/\left(|k|\cosh(\kappa)+\kappa\sinh(\kappa)\right),\\
C_{1} & = & D_{2}\cosh(\kappa)+C_{2}\sinh(\kappa)-C_{0},\\
C_{3} & = & D_{2}\cosh(\kappa)-C_{2}\sinh(\kappa).\label{eq:C3}
\end{eqnarray}

\subsection{Semi-analytical solution for the external magnetic field generated
by a pair of straight wires}

Here we extend the solution above to the external magnetic field generated
by a finite-size coil which consists of two parallel straight wires
fed with an ac current of amplitude $I_{0}$ flowing in the opposite
directions along the $z$-axis at distance $2s$ in the $x$-direction
and placed at height $h$ above the upper surface of the layer, as
shown in figure \ref{cap:sketch}. The free-space distribution of
the vector potential amplitude having only the $z$-component, which
is further scaled by $\mu_{0}I_{0},$ is governed by 
\begin{equation}
\vec{\nabla}^{2}A_{0}=-\delta(\vec{r}-h\vec{e}_{y}-s\vec{e}_{x})+\delta(\vec{r}-h\vec{e}_{y}+s\vec{e}_{x}),\label{eq:A02}
\end{equation}
where $\delta(\vec{r})$ is the Dirac delta function and $\vec{r}$
is the radius vector. The problem is solved by the Fourier transform
$\hat{A}(y;k)=\int_{-\infty}^{\infty}A(x,y)e^{\i kx}\,\d x,$ which
yields 
\begin{equation}
\hat{A}_{0}(y;k)=\i e^{-|k(y-h)|}\sin(ks)/|k|.\label{sol:A0-k}
\end{equation}
The solution for a Fourier mode with the wave number $k$ in the regions
above, below and inside the layer is given, respectively, by expressions
(\ref{sol:A-a}, \ref{sol:A-b}) and (\ref{sol:A-i}) with the coefficients
(\ref{eq:C2})\textendash (\ref{eq:C3}) containing the constant $C_{0},$
which is defined by substituting (\ref{sol:A0-k}) into (\ref{eq:C0}).
Then the spatial distribution of the complex vector potential amplitude
is given by the inverse Fourier transform $A(x,y)=\frac{1}{2\pi}\int_{-\infty}^{\infty}\hat{A}(y;k)e^{-\i kx}\,\d k,$
which is computed using the Fast Fourier Transform.

\section*{Acknowledgement}

R.L. thanks the School of Computing, Electronics and Maths at Coventry
University for funding his studentship.

\section*{References}

\bibliographystyle{elsarticle-num-names}
\bibliography{flowmeter}

\end{document}